\documentclass[12pt]{article}
\usepackage{amsfonts}
\usepackage{amssymb}
\usepackage{graphics,amsmath}

\def\hybrid{\topmargin -20pt    \oddsidemargin 0pt
        \headheight 0pt \headsep 0pt
        \textwidth 6.35in       
        \textheight 9.25in       
        \marginparwidth .875in
        \parskip 5pt plus 1pt   \jot = 1.5ex}

\hybrid

\def\baselinestretch{1.2}

\catcode`\@=11

\def\marginnote#1{}
\newcount\hour
\newcount\minute
\newtoks\amorpm
\hour=\time\divide\hour by60
\minute=\time{\multiply\hour by60 \global\advance\minute by-\hour}
\edef\standardtime{{\ifnum\hour<12 \global\amorpm={am}%
        \else\global\amorpm={pm}\advance\hour by-12 \fi
        \ifnum\hour=0 \hour=12 \fi
        \number\hour:\ifnum\minute<10 0\fi\number\minute\the\amorpm}}
\edef\militarytime{\number\hour:\ifnum\minute<10 0\fi\number\minute}

\def\draftlabel#1{{\@bsphack\if@filesw {\let\thepage\relax
   \xdef\@gtempa{\write\@auxout{\string
      \newlabel{#1}{{\@currentlabel}{\thepage}}}}}\@gtempa
   \if@nobreak \ifvmode\nobreak\fi\fi\fi\@esphack}
        \gdef\@eqnlabel{#1}}
\def\@eqnlabel{}
\def\@vacuum{}
\def\draftmarginnote#1{\marginpar{\raggedright\scriptsize\tt#1}}

\def\draft{\oddsidemargin -.5truein
        \def\@oddfoot{\sl preliminary draft \hfil
        \rm\thepage\hfil\sl\today\quad\militarytime}
        \let\@evenfoot\@oddfoot \overfullrule 3pt
        \let\label=\draftlabel
        \let\marginnote=\draftmarginnote
   \def\@eqnnum{(\theequation)\rlap{\kern\marginparsep\tt\@eqnlabel}%
\global\let\@eqnlabel\@vacuum}  }

\def\preprint{\twocolumn\sloppy\flushbottom\parindent 2em
        \leftmargini 2em\leftmarginv .5em\leftmarginvi .5em
        \oddsidemargin -.5in    \evensidemargin -.5in
        \columnsep .4in \footheight 0pt
        \textwidth 10.in        \topmargin  -.4in
        \headheight 12pt \topskip .4in
        \textheight 6.9in \footskip 0pt
        \def\@oddhead{\thepage\hfil\addtocounter{page}{1}\thepage}
        \let\@evenhead\@oddhead \def\@oddfoot{} \def\@evenfoot{} }

\def\numberbysection{\@addtoreset{equation}{section}
        \def\theequation{\thesection.\arabic{equation}}}

\def\underline#1{\relax\ifmmode\@@underline#1\else
        $\@@underline{\hbox{#1}}$\relax\fi}

\def\titlepage{\@restonecolfalse\if@twocolumn\@restonecoltrue\onecolumn
     \else \newpage \fi \thispagestyle{empty}\c@page\z@
        \def\thefootnote{\fnsymbol{footnote}} }

\def\endtitlepage{\if@restonecol\twocolumn \else \newpage \fi
        \def\thefootnote{\arabic{footnote}}
        \setcounter{footnote}{0}}  

\catcode`@=12
\relax

\def\figcap{\section*{Figure Captions\markboth
        {FIGURECAPTIONS}{FIGURECAPTIONS}}\list
        {Figure \arabic{enumi}:\hfill}{\settowidth\labelwidth{Figure
999:}
        \leftmargin\labelwidth
        \advance\leftmargin\labelsep\usecounter{enumi}}}
 \relax
\def\tablecap{\section*{Table Captions\markboth
        {TABLECAPTIONS}{TABLECAPTIONS}}\list
        {Table \arabic{enumi}:\hfill}{\settowidth\labelwidth{Table
999:}
        \leftmargin\labelwidth
        \advance\leftmargin\labelsep\usecounter{enumi}}}
 \relax
\def\reflist{\section*{References\markboth
        {REFLIST}{REFLIST}}\list
        {[\arabic{enumi}]\hfill}{\settowidth\labelwidth{[999]}
        \leftmargin\labelwidth
        \advance\leftmargin\labelsep\usecounter{enumi}}}
 \relax
\makeatletter
\newcounter{pubctr}
\def\publist{\@ifnextchar[{\@publist}{\@@publist}}
\def\@publist[#1]{\list
        {[\arabic{pubctr}]\hfill}{\settowidth\labelwidth{[999]}
        \leftmargin\labelwidth
        \advance\leftmargin\labelsep
        \@nmbrlisttrue\def\@listctr{pubctr}
        \setcounter{pubctr}{#1}\addtocounter{pubctr}{-1}}}
\def\@@publist{\list
        {[\arabic{pubctr}]\hfill}{\settowidth\labelwidth{[999]}
        \leftmargin\labelwidth
        \advance\leftmargin\labelsep
        \@nmbrlisttrue\def\@listctr{pubctr}}}
 \relax
\makeatother
\newskip\humongous \humongous=0pt plus 1000pt minus 1000pt

\newif\ifdtup

\relax

\def\be{\begin{equation}}
\def\ee{\end{equation}}
\def\ba{\begin{eqnarray}}
\def\ea{\end{eqnarray}}

\def\a{\alpha}

\def\e{\epsilon}

\def\l{\lambda}

\def\cN{{\cal N}}

\def\no{\noindent}

\def\IR{\relax{\rm I\kern-.18em R}}
\def\II{\relax{\rm 1\kern-.35em1}}

\renewcommand{\theequation}{\thesection.\arabic{equation}}
\csname @addtoreset\endcsname{equation}{section}

\def\IR{\relax{\rm I\kern-.18em R}}
\def\inv{^{\raise.15ex\hbox{${\scriptscriptstyle -}$}\kern-.05em 1}}

\begin{document}

\begin{titlepage}
\begin{center}

\vskip .5in

{\LARGE Spinning strings in $AdS_3 \times S^3$ with NS-NS flux}
\vskip 0.4in

{\bf Rafael Hern\'andez}\phantom{x} and \phantom{x}
{\bf Juan Miguel Nieto} 
\vskip 0.1in

Departamento de F\'{\i}sica Te\'orica I\\
Universidad Complutense de Madrid\\
$28040$ Madrid, Spain\\
{\footnotesize{\tt rafael.hernandez@fis.ucm.es, juanieto@ucm.es}}

\end{center}

\vskip .4in

\centerline{\bf Abstract}
\vskip .1in
\no
The sigma model describing closed strings rotating in $AdS_3 \times S^3$ is known to reduce to the one-dimensional 
Neumann-Rosochatius integrable system. In this article we show that closed spinning strings in $AdS_3 \times S^3 \times T^4$ 
in the presence of NS-NS three-form flux can be described by an extension of the Neumann-Rosochatius system. We consider 
closed strings rotating with one spin in $AdS_3$ and two different angular momenta in $S^3$. For a class of solutions with constant 
radii we find the dependence of the classical energy on the spin and the angular momenta as an expansion in the square 
of the 't Hooft coupling of the theory.

\noindent

\vskip .4in
\noindent

\end{titlepage}

\vfill
\eject

\def\baselinestretch{1.2}


\baselineskip 20pt


\section{Introduction}

Integrability has become a promising path towards a deeper understanding of the AdS/CFT correspondence. 
After the uncovering of an integrable structure underlying four-dimensional Yang-Mills theory with maximal supersymmetry~\cite{MZ}-\cite{AFS}, 
integrability has proved to be a common feature of many other examples of the correspondence. A case of special interest where integrability 
has also been discovered is the duality between string backgrounds with an $AdS_3$ factor and maximally supersymmetric two-dimensional 
conformal field theories, the AdS$_3$/CFT$_2$ correspondence. First evidences that integrability could be present in these type of backgrounds 
came from the construction of giant magnon solutions~\cite{DS}. Later on it was shown that the Green-Schwarz action of type IIB strings 
with R-R three-form flux compactified on $AdS_3 \times S^3 \times M_4$, where $M_4$ is either $T^4$ or $S^3 \times S^1$, is an integrable 
classical theory~\cite{BSZ}. This observation has boosted the analysis of the AdS$_3$/CFT$_2$ correspondence using integrability inspired 
methods~\cite{Zarembo}-\cite{BSSS2} (for a review see reference~\cite{Sfondrini}). 
Integrability has also been found recently to be a symmetry of more general $AdS_3 \times S^3 \times M_4$ backgrounds with 
a mixture of R-R and NS-NS fluxes \cite{CZ}. This discovery has preluded a series of studies of the 
AdS$_3$/CFT$_2$ correspondence with mixed fluxes using integrability~\cite{HT}-\cite{Babichenko}. 

Many of the advances in the analysis of $AdS_3 \times S^3 \times M_4$ backgrounds using integrability have been influenced 
by the developments in the study of the AdS$_5$/CFT$_4$ correspondence. In the case of closed string solutions rotating in 
$AdS_5 \times S^5$ a beautiful picture came from the identification of the corresponding lagrangian with the Neumann-Rosochatius 
integrable system~\cite{NR}. A natural question from the point of view of the AdS$_3$/CFT$_2$ correspondence 
is what is the extension of this description to backgrounds with non-vanishing fluxes. This is the problem that we will consider in this note 
for the case of closed strings rotating in $AdS_3 \times S^3 \times M_4$ with NS-NS three-form flux. 

The remaining part of the article is organized as follows. In section~2 we will consider an ansatz for a closed 
string rotating with two different angular momenta in $S^3$ and NS-NS three-form flux. The presence of flux introduces 
an additional term in the Neumann-Rosochatius system. We find a class of solutions with constant radii relying on a similar 
set of solutions in the absence of flux. In section~3 we extend the analysis to the case where the string rotates 
both in $AdS_3$ and in $S^3$. We find a class of solutions with constant radii and one spin in $AdS_3$ and two different angular 
momenta in $S^3$. We conclude in section 4 with some general remarks and a discussion on related open problems.


\section{Rotating strings in $S^3$}

\no
In this note we will be interested in closed spinning string solutions in $AdS_3 \times S^3 \times T^4$ with NS-NS three-form flux. 
The solutions that we will study will have no dynamics along the torus and thus we will not include these directions in what follows. 
The background metrics will then be
\ba
ds_{AdS_3}^2 & \!\! = \!\! & - \cosh^2 \rho \, dt^2 + d \rho^2 + \sinh^2 \rho \, d \phi^2 \ , \nonumber \\
ds_{S^3}^2 & \!\! = \!\! & d\theta^2 + \sin^2 \theta d\phi_1^2 + \cos^2 \theta d \phi_2^2 \ , 
\ea
and the NS-NS B-field will be 
\be
b_{t \phi} = q \sinh^2 \rho \ , \quad b_{\phi_1 \phi_2} = - q \cos^2 \theta \ , 
\ee
where $0 \leq q \leq 1$. 
The value $q=0$ corresponds to the case of pure R-R flux, where the theory can be formulated in terms of a Green-Schwarz coset. 
The value $q=1$ is the limit of pure NS-NS flux, and can be described by a supersymmetric WZW model. 
In the absence of flux the sigma model for closed strings rotating in $AdS_3 \times S^3$ becomes the Neumann-Rosochatius 
integrable system, which describes an oscillator on a sphere or an hyperboloid with a centrifugal potential. 
The presence of flux introduces an additional term in the lagrangian of the Neumann-Rosochatius system~\cite{NR}.  
In order to exhibit this it will be convenient to use the embedding coordinates rather than the global coordinates. 
The embedding coordinates are related to the global $AdS_3$ and $S^3$ angles by
\footnote{We will follow closely conventions and notation in \cite{NR}.} 
\ba
Y_1 + i Y_2 \! \! \! & = & \! \! \! \sinh \rho \, e^{i\phi} \ , \quad Y_3 + i Y_0 = \cosh \rho \, e^{i \, t} \ , \\
X_1 + i X_2 \! \! \! & = & \! \! \! \sin \theta \, e^{i\phi_1} \ , \quad X_3 + i X_4 = \cos \theta \, e^{i\phi_2} \ . 
\ea
In this section we will restrict the dynamics of the strings to rotation on $S^3$, so that we will take $Y_3 + i Y_0 = e^{i t}$, 
with $t=\kappa \tau$, and $Y_1=Y_2=0$. 
For the coordinates along $S^3$ we will choose an ansatz with two different angular momenta,
\be
X_1 + i X_2 = r_1 (\sigma) \, e^{i \varphi_1 (\tau, \sigma)} \ , \quad X_3 + i X_4 = r_2 (\sigma) \, e^{i \varphi_2 (\tau, \sigma)} \ ,
\label{ansatz}
\ee
where the angles will be taken to be
\be
\varphi_i (\tau,\sigma) = \omega_i \tau + \alpha_i(\sigma) \ .
\label{ansatzangle}
\ee
As we are going to consider solutions that lie on a sphere, the functions $r_i(\sigma)$ must satisfy
\be
r_1^2+r_2^2=1 \ . \label{sphere}
\ee
As we will be interested in closed strings solutions, the above functions must satisfy 
\be
r_i(\sigma + 2 \pi) = r_i(\sigma) \ , \quad \alpha_i(\sigma + 2 \pi) = \alpha_i(\sigma) + 2 \pi \bar{m}_i \ ,
\ee
with $\bar{m}_i$ some integer numbers acting as winding numbers. 
When we enter this ansatz in the Polyakov action
\be
S = \frac {\sqrt{\lambda}}{4 \pi} \int d^2 \sigma \big[ \sqrt{-h} h^{ab} G_{MN} \partial_a X^M \partial_b X^N 
- \e^{ab} B_{MN} \partial_a X^M \partial_b X^N \big] \ ,
\ee
we find the lagrangian
\be
L_{S^3} = \frac {\sqrt{\lambda}}{2 \pi} \Big[  \sum_{i=1}^2 \frac {1}{2} \big[ (r_i')^2 + r_i^2 (\a_i')^2 - r_i^2 \omega_i^2 \big] 
- \frac {\Lambda}{2} ( r_1^2 + r_2^2 - 1) + q r_2^2 \, ( \omega_1 \alpha_2' - \omega_2 \alpha_1' ) \Big] \ ,
\label{NRq}
\ee
where the prime stands for derivatives with respect to $\sigma$, $\Lambda$ is a Lagrange multiplier and we have chosen 
the conformal gauge. The first piece in (\ref{NRq}) is the Neumann-Rosochatius integrable system \cite{NR}. The presence 
of the non-vanishing flux introduces the last term in the lagrangian. \footnote{Note that in the 
WZW model limit $q=1$
the lagrangian simplifies greatly because we can complete squares. 
We will find further evidence on this simplification below.}

We will now write the equations of motion. The lagrangian is cyclic on the variables~$\alpha_i$. Therefore we easily conclude that 
\be
\alpha_i' = \frac {v_i + q r_2^2 \epsilon_{ij} \omega_j}{r_i^2} \ , \quad i = 1,2 \ ,
\label{alphaprime}
\ee
where $v_i$ are some integrals of motion and $\epsilon_{12}=+1$ (we assume summation on $j$). 
The variation of the lagrangian with respect to the radial coordinates gives us
\begin{align}
r_1''&=-r_1 \omega_1^2 +r_1 \alpha_1^{'2}-\Lambda r_1 \ , \label{r1} \\
r_2''&=-r_2 \omega_2^2 +r_2 \alpha_2^{'2}-\Lambda r_2 + 2 q r_2 ( \omega_1 \alpha_2' - \omega_2 \alpha_1' )\ .
\label{r2}
\end{align}
To these equations we have to add the Virasoro constraints,
\begin{align}
&\sum_{i=1}^2 \big( r_i^{'2} +r_i^2 ( \alpha_i^{'2}+ \omega_i^2 ) \big)=\kappa^2 \ , \\
&\sum_{i=1}^2 r_i^2 \omega_i \alpha_i'=0 \ , \label{virasoro1} 
\end{align}
In terms of the integrals $v_i$ the second Virasoro constraint can be rewritten as
\begin{equation}
\omega_1 v_1+\omega_2 v_2=0 \ .
\end{equation}
The energy and the angular momenta of the string are given by
\begin{align}
E & = \sqrt{\lambda} \, \kappa \ , \label{E} \\
J_1 & = \sqrt{\lambda} \int_0^{2\pi} {\frac {d\sigma}{2 \pi} \left( r_1^2 \omega_1 - q r_2^2 \alpha_2' \right)} \ , \label{J1} \\
J_2 & = \sqrt{\lambda} \int_0^{2 \pi} {\frac {d\sigma}{2 \pi} \left( r_2^2 \omega_2 + q r_2^2 \alpha_1' \right)} \label{J2} \ .
\end{align}

\subsection{Constant radii solutions}

A simple solution to the equations of motion can be obtained if we take the radii $r_i$ to be some constants, $r_i=a_i$. 
In this case the derivatives of the angles also become constant and thus
\be
\alpha_i = \bar{m}_i \sigma + \a_{0i} \ ,
\ee
where the windings become
\be
\bar{m}_i \equiv \frac {v_i + q a_2^2 \epsilon_{ij} \omega_j}{a_i^2} \ .
\ee
The integration constants $\a_{0i}$ can be set to zero through a rotation, and the constants $\bar{m}_i$ must be integers 
in order to satisfy the closed string periodicity condition. The equations of motion for $r_i$ reduce now to
\begin{align}
& \omega_1^2 - \bar{m}_1^2 + \Lambda = 0 \label{o1} \ , \\
& \omega_2^2 -\bar{m}_2^{2} - 2 q ( \omega_1 \bar{m}_2 - \omega_2 \bar{m}_1 )+ \Lambda = 0 \label{o2} \ ,
\end{align}
and thus we conclude that the Lagrange multiplier $\Lambda$ is constant on this solution. The Virasoro constraints can then be written as
\begin{align}	
& \sum_{i=1}^2 a_i^2 \left( \bar{m}_i^{2}+\omega_i^2 \right) = \kappa^2 \ , \label{energy} \\
& \bar{m}_1 J_1 + \bar{m}_2 J_2 = 0 \ . \label{virasoro2}
\end{align}
We will now find the energy as a function of the angular momenta and the integer numbers $\bar{m}_i$. In order to do this we will first use 
equations (\ref{sphere}) and (\ref{virasoro1}) to write the radii as functions of $\omega_i$ and $\bar{m}_i$,
\be
a_1^2 = \frac{\omega_2 \bar{m}_2}{\omega_2 \bar{m}_2-\omega_1 \bar{m}_1} \ , \quad 
a_2^2 = \frac{\omega_1 \bar{m}_1}{\omega_1 \bar{m}_1 - \omega_2 \bar{m}_2} \ .
\ee
With these relations at hand and the definitions (\ref{E})-(\ref{J2}), together with (\ref{energy}), we find
\be
E^2 = \frac{(J_1+ \sqrt{\lambda} q a_2^2 \bar{m}_2)^2}{a_1^2} +\frac{(J_2- \sqrt{\lambda} q a_2^2 \bar{m}_1)^2}{a_2^2} 
+ \lambda \left( a_1^2 \bar{m}_1^2 + a_2^2 \bar{m}_2^2 \right) \ ,
\ee
or after some immediate algebra,
\begin{align}
E^2 & = (J_1+J_2)^2 +J_1 J_2 \frac{(1 - \hbox{w})^2}{\hbox{w}} -2 \sqrt{\lambda} q \bar{m}_1 ( J_1 \hbox{w} + J_2 ) \notag \\
& + \lambda \left( \bar{m}_ 1 \bar{m}_2 - q^2 \bar{m}_1^2 \hbox{w} \right) \frac{\bar{m}_1 - \bar{m}_2 \hbox{w}}{\bar{m}_2 - \bar{m}_1 \hbox{w}} \ ,
\label{EJ1J2}
\end{align}
where we have made use of (\ref{virasoro2}) and we have introduced $\hbox{w} \equiv \omega_1/\omega_2$. 
Now we need to write the ratio $\hbox{w}$ as a function of the windings $\bar{m}_i$ and the angular momenta $J_i$. 
This can be done by adding equations (\ref{J1}) and (\ref{J2}), subtracting equation (\ref{o2}) from (\ref{o1}), 
and solving the resulting system of equations,
\begin{align}
& \big[ \bar{m}_1 J- \sqrt{\lambda} q \bar{m}_1 (\bar{m}_1 -\bar{m}_2) \big] \hbox{w} - \bar{m}_2  J - \sqrt{\lambda} (\bar{m}_1 -\bar{m}_2)\omega_1 = 0 \ , 
\label{J1J2} \\
& \omega_1^2 -\bar{m}_1^2-\frac{\omega_1^2}{\hbox{w}^2} +\bar{m}_2^2 +2 q \bar{m}_2 \omega_1 -2 q \bar{m}_1 \frac{\omega_1}{\hbox{w}} = 0 \ ,
\end{align}
where $J \equiv J_1+J_2$ is the total angular momentum. When we eliminate $\omega_1$ in these expressions we are left 
with a quartic equation
\begin{align}
& (\bar{m}_1 \hbox{w} - \bar{m}_2 )^2 \Big[ 1 - \Big( 1 - \frac {\sqrt{\lambda}}{J} q (\bar{m}_1 -\bar{m}_2 ) \Big)^2 
\hbox{w}^2 \Big] \nonumber \\
& + \frac {\lambda}{J^2} \hbox{w}^2 (\bar{m}_1 + \bar{m}_2 ) (\bar{m}_1 - \bar{m}_2 )^3 (1- q^2) = 0 \ .
\label{quartic}
\end{align}
Rather than trying to solve this equation explicitly, we can write the solution as a power series expansion in large $J/\sqrt{\lambda}$. 
\footnote{Alternatively we can solve equation (\ref{quartic}) around the WZW limiting point $q=1$ to get
\begin{align*}
\omega_1 & = \frac {J}{\sqrt{\lambda}} + \frac {\sqrt{\lambda} \bar{m}_1 (\bar{m}_1 + \bar{m}_2) J}{(J + \sqrt{\lambda} \bar{m}_2)^2} (1-q) +  \dots \ , \\
\omega_2 & = \frac {J}{\sqrt{\lambda}} - (\bar{m}_1 -\bar{m}_2) 
+ \frac {(J^2 - \lambda \bar{m}_1 \bar{m}_2) (\bar{m}_1 - \bar{m}_2) + \sqrt{\lambda} \bar{m}_2 J (3 \bar{m}_1 - \bar{m}_2)}{(J + \sqrt{\lambda} \bar{m}_2)^2} (1-q) + \dots  \ , 
\end{align*}
which reduce to (\ref{w1}) and (\ref{w2}) in the limit where $J/\sqrt{\lambda}$ is large. 
}
Out of the four different solutions to (\ref{quartic}), the only one with a well-defined expansion is 
\be
\hbox{w} =1 + \frac {\sqrt{\lambda}}{J} q ( \bar{m}_1-\bar{m}_2)  
+ \frac {\lambda}{2J^2} ( \bar{m}_1-\bar{m}_2) \big( \bar{m}_1+\bar{m}_2 +q^2 ( \bar{m}_1 - 3 \bar{m}_2) \big)  + \cdots 
\label{w}
\ee
which implies that
\begin{align}
\omega_1 & = \frac {J}{\sqrt{\lambda}} + \frac {\sqrt{\lambda}}{2J} \bar{m}_1 (\bar{m}_1 +\bar{m}_2) (1-q^2)  
\Big[ 1 - 2 \frac {\sqrt{\lambda}}{J} q \bar{m}_2 + \dots \Big] \ , \label{w1} \\
\omega_2 & = \frac {J}{\sqrt{\lambda}} - q (\bar{m}_1 -\bar{m}_2) + \frac {\sqrt{\lambda}}{2J} \bar{m}_2 (\bar{m}_1 + \bar{m}_2) (1-q^2) 
\Big[ 1 - \frac {\sqrt{\lambda}}{J} q (\bar{m}_1 + \bar{m}_2) + \dots \Big] \ . \label{w2}
\end{align}
Note that the ${\cal O}(\sqrt{\lambda}/J)$ terms and the subsequent corrections in (\ref{w1}) and (\ref{w2}) are dressed 
with a common factor of $\bar{m}_1 + \bar{m}_2$ that vanishes for equal angular momenta. 
We can easily prove the existence of this factor if we set $\bar{m}_1 = - \bar{m}_2 \equiv m$ in equation (\ref{quartic}), which reduces to
\be
(1 + \hbox{w})^2 \big[ (J - 2 \sqrt{\lambda} q m )^2 \hbox{w}^2 - J^2 \big] = 0 \ ,
\ee
whose only well-defined solution is
\be
\hbox{w} = \frac {J}{J - 2 \sqrt{\lambda} q m} \ ,
\label{wexact}
\ee
and therefore we can calculate the frequencies $\omega_1$ and $\omega_2$ exactly, 
\be
\omega_1 = \frac {J}{\sqrt{\lambda}} \ , \quad 
\omega_2 = \frac {J}{\sqrt{\lambda}} - 2 q m \ . 
\ee
An identical reasoning can be employed to prove the existence of the global factor $1-q^2$.

If we substitute the value of $\hbox{w}$ in equation (\ref{w}) in relation (\ref{EJ1J2}) we find
\be
E^2 = J^2  - 2 \sqrt{\lambda} q \bar{m}_1 J + \frac {\lambda}{J} \big[ (\bar{m}_1^2 J_1+\bar{m}_2^2 J_2) (1-q^2) 
+ q^2 \bar{m}_1^2 J \big] + \cdots 
\label{qdispersion}
\ee
When the flux vanishes this expression becomes the expansion for the energy in the Neumann-Rosochatius system 
describing closed string solutions rotating with two different angular momenta \cite{NR}. 
We must note that the subleading terms not included in (\ref{qdispersion}) contain a common factor of $\bar{m}_1 + \bar{m}_2$. 
Therefore if we look at the particular case of $\bar{m}_1= - \bar{m}_2$ relation (\ref{qdispersion}) simplifies to 
\be
E^2 = J^2 - 2 \sqrt{\lambda} q m J + \lambda m^2 \ .
\label{HSTdispersion}
\ee
This is the expression for the energy in the case of circular string solutions with two equal angular momenta found in \cite{HST}. 
We stress that relation (\ref{HSTdispersion}) is an exact result because the ratio $\hbox{w}$ is given by equation (\ref{wexact}). 

In a similar way if we focus on the case of pure NS-NS flux, where $q=1$, the energy can also be found exactly,
\begin{align}
E^2 & = \frac {1}{J} \Big[ J^3 + (\sqrt{\lambda} \bar{m}_1 - J_2) J_1^2 - (\sqrt{\lambda} \bar{m}_1 + J_1) J_2^2 - 2 \sqrt{\lambda} \bar{m}_1 J^2 + \lambda \bar{m}_1^2 J \nonumber \\
& - \sqrt{\lambda} (\bar{m}_1 - \bar{m}_2) J_1 J_2 - \frac { (\sqrt{\lambda} \bar{m}_1 + J_1)(\sqrt{\lambda} \bar{m}_1 - J_2) J^2}{J- \sqrt{\lambda} (\bar{m}_1 - \bar{m}_2)} 
\Big] \ ,
\end{align}
which reduces to 
\be
E = J - \sqrt{\l} \bar{m}_1 
\ee
when we write the angular momenta $J_1$ and $J_2$ in terms of the total momentum $J$.


\section{Rotating strings in $AdS_3  \times S^3$}

We will now extend the analysis in the previous section to the case where the string can rotate both in $AdS_3$ and $S^3$, 
again with no dynamics along $T^4$. The string solutions that we will consider will therefore have one spin $S$ in $AdS_3$ 
and two angular momenta $J_1$ and $J_2$ in $S^3$. We can describe these configurations 
with the ansatz (\ref{ansatz})--(\ref{ansatzangle}), together with 
\be
Y_3 + i Y_0 = z_0 (\sigma) \, e^{i \phi_0 (\tau, \sigma)} \ , \quad Y_1 + i Y_2 = z_1 (\sigma) \, e^{i \phi_1 (\tau, \sigma)} \ ,
\label{ansatzAdS}
\ee
where the angles are
\be
\phi_a (\tau,\sigma) = w_a \tau + \beta_a (\sigma) \ , 
\ee
together with the periodicity conditions
\be
z_a(\sigma + 2 \pi) = z_a (\sigma) \ , \quad \beta_a(\sigma + 2 \pi) = \beta_a(\sigma) + 2 \pi \bar{k}_a \ ,
\ee
with $a=0,1$. Note however that the time direction has to be single-valued so we need to exclude windings 
along the time coordinate. Therefore we must take $\bar{k}_0=0$. When we substitute this ansatz in the Polyakov action 
in the conformal gauge we obtain again lagrangian (\ref{NRq}) for rotation in the $S^3$ piece, together with the contribution 
from $AdS_3$,
\begin{equation}
L_{AdS_3} = \frac{\sqrt{\lambda}}{4\pi} \Big[ g^{ab} \left( z'_a z'_b + z_a z_a \beta_b'^2 - z_a z_a w_b^2 \right) 
-\frac {\tilde{\Lambda}}{2} \left( g^{ab} z_a z_b +1 \right) - 2 q z_1^2 ( w_0 \beta '_1 - w_1 \beta '_0 ) \Big] \ ,
\end{equation}
where we have chosen $g=\hbox{diag}(-1,1)$ and $\tilde{\Lambda}$ is a Lagrange multiplier. We will now write the equations of motion.  
As the pieces of the lagrangian describing motion along $AdS_3$ and $S^3$ are decoupled the equations of motion for $r_i$ and $\alpha_i$ 
are given directly by expressions (\ref{alphaprime})--(\ref{r2}). In a similar way, the equations of motion for $z_a$ are
\begin{align}
z''_0 &=z_0 \beta^{'2}_0 -z_0 w_0^2 -\tilde{\Lambda} z_0 \ , \\
z''_1 &=z_1 \beta^{'2}_1 -z_1 w_1^2 -\tilde{\Lambda} z_1 -2 q z_1 ( w_0 \beta'_1 - w_1 \beta'_0 ) \ ,
\end{align}
and the equations for the angles are 
\be
\beta '_a = \frac{u_a + q z_1^2 \epsilon_{ab} w_b}{g^{aa} z_a^2} \ , \\
\ee
where $u_a$ are some integration constants. To these equations we need to add the constraint 
\be
- z_0^2 + z_1^2 = - 1 \ ,
\label{AdSconstraint}
\ee
together with the Virasoro constraints, which are responsible for the coupling between the $AdS_3$ and the $S^3$ systems, 
\begin{align}
& z^{'2}_0 + z_0^2 (\beta ^{'2}_0 + \kappa^2 ) = z^{'2}_1 + z_1^2 (\beta ^{'2}_1 + w_1^2)
+ \sum_{i=1}^2 \big( r^{'2}_i + r_i^2 (\alpha^{'2}_i + \omega^2_i ) \big)  \ , \\
& z_1^2 w_1 \beta '_1 + \sum_{i=1}^2 r_i^2 \omega _i \alpha '_i  = z_0^2 \kappa \beta '_0 \ ,
\end{align}
where we have set $w_0 \equiv \kappa$. The spin and the energy in this case are given by
\begin{align}
E & = \sqrt{\lambda} \int_0^{2\pi} {\frac{d\sigma}{2\pi} (z_0^2 \kappa - q z_1^2 \beta'_1)} \ , \label{EAdS} \\
S & = \sqrt{\lambda} \int_0^{2\pi} {\frac{d\sigma}{2\pi} 
(z_1^2 w_1 -q z_1^2 \beta'_0)} \ , \label{SAdS}
\end{align}
and the angular momenta are defined again as in equations (\ref{J1}) and (\ref{J2}).

\subsection{Constant radii solutions}

As before a simple solution to these equations can be found when the string radii are taken as constant, 
$r_i=a_i$ and $z_a=b_a$. In this case the periodicity condition on $\beta_0$ and the fact that the time coordinate is single-valued implies 
\be
\beta'_0=0 \ .
\ee
Furthermore the angles can be easily integrated again,
\be
\beta'_1= \bar{k} \ , \quad \alpha'_i=\bar{m}_i \ , \quad i=1,2 \ ,
\ee
and thus the equations of motion reduce to
\begin{align}
& w_1^2 -\bar{k}^2 - \kappa^2 + 2 q \kappa \bar{k} = 0 \ , \label{eom1} \\ 
& (\omega_2^2 -\omega_1^2)-(\bar{m}_2^2 -\bar{m}_1^2) - 2q ( \omega_1 \bar{m}_2 - \omega_2 \bar{m}_1 ) = 0 \label{eom2} \ .
\end{align}
The Virasoro constraints become then
\begin{align}
& b_1^2 (w_1^2 +\bar{k}^{2}) + \sum_{i=1}^2 a_i^2 (\omega^2_i + \bar{m}^{2}_i) = b_0^2 \kappa^2 \ , \\
& \bar{k} S + \bar{m}_1 J_1 + \bar{m}_2 J_2 = 0 \ .
\end{align}
Using the definitions of the energy and the spin, equations (\ref{EAdS}) and (\ref{SAdS}), together with the constraint (\ref{AdSconstraint}), 
we can write
\be
E_{\pm} =  \sqrt{\lambda} \, \kappa \pm \frac{S (\kappa - q \bar{k})}{\sqrt{\kappa^2 + \bar{k}^2 - 2 q \bar{k} \kappa }} \ .
\label{ESJ1J2}
\ee
The plus sign corresponds to the case where $\kappa$ and $w_1$ are chosen to have equal signs, while the minus sign corresponds 
to the choice of opposite signs. 
We can use now this expression to write the energy as a function of the spin, the two angular momenta 
and the winding numbers $\bar{k}$ and $\bar{m}_i$. As in the previous section we can take the second 
Virasoro constraint together with the condition that $a_1^2+a_2^2=1$ to find that 
\be
a_1^2 = \frac{ \bar{k} S + \sqrt{\lambda} \omega_2 \bar{m}_2}{\sqrt{\lambda} (\omega_2 \bar{m}_2 - \omega_1 \bar{m}_1)} \ , \quad 
a_2^2 = \frac{ \bar{k} S +\sqrt{\lambda} \omega_1 \bar{m}_1}{\sqrt{\lambda} (\omega_1 \bar{m}_1 - \omega_2 \bar{m}_2)} \ . 
\label{a1a2}
\ee
Taking these relations into account when adding the angular momenta (\ref{J1}) and (\ref{J2}) we find
a relation between the frequencies $\omega_1$ and $\omega_2$,
\begin{align}
& \left[ \bar{k} S + \bar{m}_1 J -\sqrt{\lambda} q \bar{m}_1 (\bar{m}_1 - \bar{m}_2) \right] 
\frac{\omega_1}{\omega_2} - ( \bar{k} S + \bar{m}_2 J ) \nonumber \\ 
& - \sqrt{\lambda} ( \bar{m}_1 - \bar{m}_2) \, \omega_1 - \frac{q \bar{k} S (\bar{m}_1 - \bar{m}_2)}{\omega_2} = 0 \ ,
\label{w1w2}
\end{align}
which extends expression (\ref{J1J2}) to the case of spin in $AdS_3$. Combining now equation (\ref{eom2}) with (\ref{w1w2}) 
we can solve for $\omega_1$. The result is again a quartic equation,
\begin{align}
& \Big[(\omega_1 +q \bar{m}_2)^2 -(\bar{m}_1^2-\bar{m}_2^2) (1-q^2) \Big] \Big[ \lambda (\bar{m}_1-\bar{m}_2)\omega_1^2
+ 2\sqrt{\lambda} (\bar{m}_2 J+ \bar{k} S) \omega_1 \notag \\
& - \big( ( \bar{m}_1 + \bar{m}_2 ) J + 2 \bar{k} S \big) J \Big]
- ( \bar{m}_1 + \bar{m}_2 ) ( \bar{m}_1 J + \bar{k} S )^2 (1-q^2) = 0 \ .
\label{quarticAdS}
\end{align}
Once we have found the solution to this equation, we can read $\omega_2$ from (\ref{w1w2}) and 
use then the first Virasoro constrain to calculate $\kappa$. But before writing the resulting equation let us first take into account that
\be
b_1^2 w_1^2 +b_1^2 \bar{k}^2 -b_0^2 \kappa^2= b_1^2 (2\bar{k}^2 -2q \kappa \bar{k})- \kappa^2 
= \frac{ 2\bar{k} S (\bar{k}-q \kappa)}{\sqrt{\lambda (\kappa^2 +\bar{k}^2 -2q \bar{k} \kappa )}} - \kappa^2 \ ,
\ee
where we have made use of (\ref{eom1}). The Virasoro constraint becomes thus a sixth-grade equation for $\kappa$,
\begin{equation}
\frac{4\bar{k}^2 S^2 (\bar{k}-q\kappa)^2}{\lambda (\kappa^2 + \bar{k}^2 -2q\bar{k} \kappa)}=\big( \kappa^2 - a_1^2 (\omega_1^2 + \bar{m}_1^2 ) 
- a_2^2 ( \omega_2^2 + \bar{m}_2^2) \big)^2 \ .
\label{kappaequation}
\end{equation}
The solution to this equation provides $\kappa$, and thus the energy, as a function of the spin, the angular momenta, 
and the winding numbers $\bar{k}$ and $\bar{m}_i$. However equations (\ref{quarticAdS}) and (\ref{kappaequation}) 
are difficult to solve exactly. As in the previous section, instead of trying to find an exact solution we can write the solution 
in the limit $J_i/\sqrt{\lambda}  \sim S/\sqrt{\lambda} \gg 1$. Out of the four different solutions to (\ref{quarticAdS}), 
the only one with a well-defined limit is 
\be
\omega_1 = \frac {J}{\sqrt{\lambda}} + \frac{\sqrt{\lambda}}{2J^2} (\bar{m}_1 + \bar{m}_2) (\bar{m}_1 J + \bar{k} S ) (1-q^2) 
\Big[ 1 - \frac {\sqrt{\lambda}}{J} q \bar{m}_2 + \cdots \Big] \ .
\label{w1AdS}
\ee
Using now relation (\ref{w1w2}) we find 
\footnote{Note that as in the case of rotation just in the sphere the ${\cal O}(\sqrt{\lambda}/J)$ terms and the subsequent corrections 
in the expansions for $\omega_1$ and $\omega_2$ are again proportional to $\bar{m}_1 + \bar{m}_2$. We can prove the existence 
of this factor as in the previous section by setting $\bar{m}_1 = - \bar{m}_2$ in equation (\ref{quarticAdS}).}
\begin{align}
\omega_2 & = \frac {J}{\sqrt{\lambda}} - q ( \bar{m}_1 - \bar{m}_2) + \frac {\sqrt{\lambda}}{2J^2} (\bar{m}_1 + \bar{m}_2) (1-q^2) \nonumber \\
& \times \Big[ \bar{m}_2 J + \bar{k} S 
- \frac {\sqrt{\lambda}}{J} q \bar{m}_2 (\bar{m}_1 J + \bar{m}_2 J + 2 \bar{k} S ) + \cdots \Big] \ .
\label{w2AdS}
\end{align}
Next we can calculate the radii $a_1$ and $a_2$ using (\ref{a1a2}), and solve equation (\ref{kappaequation}) to get
\begin{align}
\kappa_+ & = \frac {J}{\sqrt{\lambda}} - q \left( \bar{m}_1 + 2 \frac{ \bar{k} S}{J} \right) 
+ \frac{\sqrt{\lambda}}{2 J^2}(\bar{m}_1^2 J_1+\bar{m}_2^2 J_2 + 2 \bar{k}^2 S) (1-q^2) 
\notag \\
& - \frac{\sqrt{\lambda}}{J^3} 2 q^2  \bar{k} S ( \bar{m}_1 J + \bar{k} S ) + \cdots \\
\kappa_- & = \frac {J}{\sqrt{\lambda}} - q \bar{m}_1 
+ \frac{\sqrt{\lambda}}{2 J^2}(\bar{m}_1^2 J_1+\bar{m}_2^2 J_2 - 2 \bar{k}^2 S) (1-q^2)  + \cdots \\
\end{align}
where as in equation (\ref{ESJ1J2}) the plus or minus subindices refer respectively to the cases where $\kappa$ and $w_1$ are chosen 
with identical or opposite signs. These expressions can now be substituted in relation (\ref{ESJ1J2}) to obtain
\begin{align}
E_+ & = J + S - \sqrt{\lambda} q \left( \bar{m}_1 + 2 \frac{ \bar{k} S}{J} \right) + \frac{\lambda}{2 J^2}(\bar{m}_1^2 J_1+\bar{m}_2^2 J_2 + \bar{k}^2 S) (1-q^2)  \notag \\
& - \frac{\lambda}{J^3} 2 q^2 \bar{k} S (\bar{m}_1 J+\bar{k} S) + \cdots \ ,\\
E_- & = J - S - \sqrt{\lambda} q \bar{m}_1 + \frac{\lambda}{2 J^2}(\bar{m}_1^2 J_1+\bar{m}_2^2 J_2 - \bar{k}^2 S) (1-q^2) + \cdots 
\end{align}
In the absence of flux the expression for $E_+$ reduces to the expansion for the energy in the Neumann-Rosochatius system for a closed circular 
string of constant radius rotating with one spin in $AdS_3$ and two different angular momenta in $S^3$ \cite{NR}.

As in the previous section, we can now consider the limit of pure NS-NS flux. In this case the above expressions simplify greatly, and we get
\begin{align}
E_+ & = S + \sqrt{ \big(J - \sqrt{\lambda} \bar{m}_1 \big)^2 - 4 \sqrt{\lambda} \bar{k} S} \ , \\
E_- & = J - S - \sqrt{\lambda} \bar{m}_1 \ .
\end{align}


\section{Concluding remarks}

In this letter we have studied closed string solutions rotating in $AdS_3 \times S^3 \times T^4$ with NS-NS three-form flux. 
The corresponding string sigma model reduces to the Neumann-Rosochatius integrable system with an additional 
contribution coming from the non-vanishing flux term. We have considered the cases where the string can rotate either 
in $S^3$ with two different angular momenta, or in $AdS_3 \times S^3$ with one spin and two different angular momenta. 
The equations of motion can be easily integrated either as a power series in $J/\sqrt{\lambda}$ or as a power series around 
the pure NS-NS point $q=1$ for the case of constant radii strings. We have found the classical energy in terms of the conserved quantities 
and the parameter governing the strength of the NS-NS flux. 

There are many natural extensions of our analysis in this note. An immediate one is the choice of an ansatz where the worldsheet 
coordinates $\tau$ and $\sigma$ are exchanged. This kind of ansatz, where the radial coordinates depend on the time variable, 
corresponds to the pulsating string solutions considered in~\cite{GKP,Minahanpulsating}. A similar solution can indeed be readily constructed 
also in the presence of non-vanishing three-form flux. An equally straightforward continuation of our analysis is the analysis 
of more general solutions. It would be for instance very interesting to find elliptic solutions with non-vanishing flux, 
following the analysis in the case of the Neumann-Rosochatius system~\cite{NR}.

Another important question is the study of the spectrum of small quadratic fluctuations~\cite{FT,NR} around the circular 
solutions that we have constructed in this note. This problem is a necessary step in order to determine the conditions 
of stability of our solutions and to find the spectrum of excited string states. It would also be very interesting to extend the analysis in this note to deformations 
of the $AdS_3$ backgrounds. An appealing case is that of the $\eta$-deformation of $AdS_5 \times S^5$ \cite{DMV}, which has been 
recently shown to lead to an integrable extension of the Neumann-Rosochatius system \cite{AM}. However the $\eta$-deformed Neumann-Rosochatius system 
is much more involved than the one that we have obtained in this note because the deformation is obtained by breaking the isometries of the metric down 
to the Cartan algebra. Currently only the metric and the NS-NS flux are known for the $\eta$-deformation, but if a complete solution was constructed one could 
expect to be able to introduce a spinning string ansatz with mixed fluxes, depending on the $q$ and $\eta$ parameters.


\vspace{8mm}

\centerline{\bf Acknowledgments}

\vspace{2mm}

\no
The work of R.~H. is supported by MICINN through a Ram\'on y Cajal contract and grant FPA2011-24568, 
and by BSCH-UCM through grant GR58/08-910770. J.~M.~N. wishes to thank the Instituto de F\'{\i}sica Te\'orica UAM-CSIC 
for kind hospitality during this work. 


\end{document}